\documentclass[twocolumn,10pt]{article}
\usepackage{epsfig}
\usefont{T1}{ptm}{m}{n}
\begin{document}

\def\rv{{\bf r}}
\def\vv{{\bf v}}
\def\av{{\bf a}}
\def\bv{{\bf b}}
\def\lv{{\bf l}}
\def\ev{{\bf e}}
\def\dv{{\bf d}}
\def\pv{{\bf p}}
\def\qv{{\bf q}}
\def\kv{{\bf k}}
\def\jv{{\bf j}}
\def\mv{{\bf m}}
\def\Ev{{\bf E}}
\def\Bv{{\bf B}}
\def\Dv{{\bf D}}
\def\Hv{{\bf H}}
\def\Pv{{\bf P}}
\def\Mv{{\bf M}}
\def\Av{{\bf A}}
\def\Sv{{\bf S}}
\def\Jv{{\bf J}}
\def\fv{{\bf f}}
\def\nv{\hat{\bf n}}
\def\grad{{\bf \nabla}}
\def\div{{\bf \nabla \cdot}}
\def\rot{{\bf \nabla} \times}
\def\magmom{m}
\def\antidual{^\star}
\def\dual{^\star}

%
\title{ 
DYNAMICS OF A MAGNETIC MONOPOLE IN MATTER, \\
MAXWELL EQUATIONS IN DYONIC MATTER \\
AND DETECTION OF ELECTRIC DIPOLE MOMENTS
}
 
\author{
{\bf 
X. Artru$^a$, D. Fayolle$^b$
} \\ 
{\small $^a$ Institut de Physique Nucl\'eaire de Lyon, 
IN2P3-CNRS \& Universit\'e Claude Bernard, France} \\
{\small $^b$ Laboratoire de Physique Corpusculaire de Clermont, 
IN2P3-CNRS \& Universit\'e Blaise Pascal, France}
}
\date{
{\begin{flushleft} \normalsize
For a monopole, the analogue of the Lorentz equation in matter
is shown to be $\fv = g \ (\Hv - \vv \times \Dv)$. 
Dual-symmetric Maxwell equations,
for matter containing hidden magnetic charges 
in addition to electric ones, are given.
They apply as well to ordinary matter if the particles
possess T-violating electric dipole moments. 
Two schemes of experiments for the detection of
such moments in macroscopic pieces of matter are proposed.\\
~~PACS: 14.80.Hv ~ 03.50.De ~ 11.30.Er 
\end{flushleft}
}}

\maketitle



\noindent
{\bf 1. INTRODUCTION} 

The question of which classical macroscopic fields exert a force 
on a magnetic monopole of charge $g$ in matter is still controversial [1].
For the static force, the formula
$$
\fv = g \, \Hv
\,,
\eqno(1)
$$
instead of $\fv = g \, \Bv$, is generally accepted.
However, for the velocity dependent force, there is no consensus between 
$\fv = - g \, \vv \times \Ev$ and $\fv = - g \, \vv \times \Dv$
(we use rationalized equations with $c=\varepsilon_0=\mu_0=1$).
A more general problem is to generalize the macroscopic Maxwell equations
to the dual-symmetric matter.
The atoms or molecules of such a matter would be made not only
of electrically, but also of magnetically charged particles.
Thus they can possess
\begin{itemize}
\item electric dipole moments coming from decentered electric charges
as well as spinning magnetic charges
\item magnetic moments comming from spinning electric charges
as well as decentered magnetic charges.
\end{itemize}

\noindent
After a rederivation and a discussion of Eq.(1), we will present below
a consistent solution for the velocity-dependent force 
and the dual-symmetric Maxwell equations in matter,
using simple physical arguments. 
We will consider only isotropic matter and assume that its
electric and magnetic polarizations $\Pv$ and $\Mv$ 
are linear in $\Dv$ and $\Bv$ (or $\Ev$ and $\Hv$).
It will appear that our equations can also take into account
the electric dipole moments (e.d.m.) of the ordinary fermions 
generated by T-violating interactions, and we will propose
two kinds of possible measurements of the e.d.m. in macroscopic 
matter.

\medskip
\noindent
{\bf 2. STATIC FORCE ON A MONOPOLE IN MATTER.}

If the force acting on a monopole in matter were $\fv = g \, \Bv$,
a monopole following a closed magnetic line of a permanent
magnet could gain energy at each turn, providing a perpetual motion
of the first species. This is an argument for chosing $\fv = g \, \Hv$,
the curl of which is zero for a static system.

One might object that the monopole can gain energy at each turn
at the expense of the magnetic energy stored by the magnet 
and will eventually erase the magnetization of the metal. 
This is indeed what happens when a monopole is circulating through
a superconducting loop : the varying flux of the monopole
field through the loop produces a counter-electromotive force
which damps the supercurrent. 
However, in the case of a ferroelectric annulus,
the magnetized state has the lowest energy
and the annulus cannot yield any energy to the monopole.

Another argument for (1) comes from the (gedanken) following experiment~:
Let us measure the force on a magnetic charge immersed in a ferrofluid. 
The latter is a practical realization of a liquid magnetizable\ matter.
No static frictional force can perturb the measurement.
We protect the monopole from the fluid by a waterproof box. 
This should not change the result~; 
anyway the physical monopole is probably dressed 
by a swarm of ordinary particles. 
In the absence of the monopole, we denote by $\Bv \equiv \mu \Hv$ the field 
outside the box and by $\Bv_{box}$ = $\Hv_{box}$ the field inside the box.
The fields coming from the monopole will be denoted by a prime.
Let us consider two shapes of box (Fig.1)~:

\begin{figure}
\centering\epsfig{file=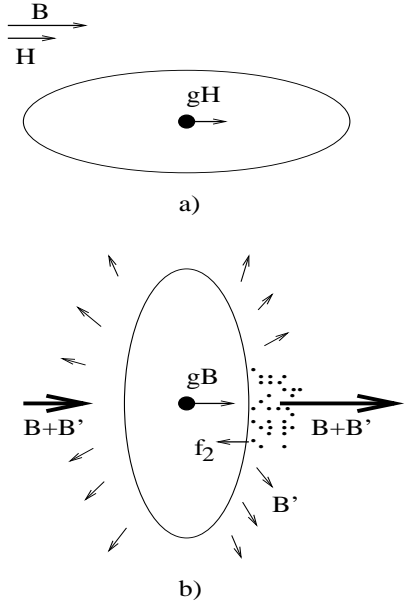,width=5.4cm,height=7.9cm}
\caption{Force acting on a monopole in a ferrofluid. 
(a) elongated box parallel to the field
(b) flattened box perpendicular to the field}
\end{figure}

\noindent
a) the box is elongated parallel to $\Bv$. 
Then $\Bv_{box}$ = $\Hv$
and the measured force is $\fv = g \, \Hv$.

\noindent
b) the box is flattened perpendicular to $\Bv$. 
Then $\Bv_{box}$ = $\Bv$.
The force acting on the pole is $\fv_1 = g \, \Bv$, 
which is different from case a).
On the other hand, in front of the box the total field 
$\Bv^{tot} = \Bv + \Bv'$ is larger than behind.
The magnetic grains of the ferrofluid are therefore attracted 
toward this region and build a hydrostatic pressure 
which pushes the box backwards.
Quantitatively, the force acting on one grain of magnetic moment 
$\vec\magmom$ in the nonuniform field $\Bv^{tot}$ is
$$
f_i = \magmom_j \ \partial_j B^{tot}_i = \magmom_j \ \partial_i B^{tot}_j
\,,
\eqno(2)
$$
since $\rot \Bv^{tot} = 0$ for a static system.
Repeated indices are summed over.
The resulting macroscopic force by unit volume is
$$
{dF_i \over d^3\rv} = M^{tot}_j \ \partial_i B^{tot}_j
= {\chi\over2} \ \partial_i \, (\Bv^{tot} \cdot \Bv^{tot})
$$
where $\Mv^{tot} = \Mv+\Mv' = \chi (\Bv+\Bv')$ 
is the magnetization density and $\chi={\mu-1\over\mu}$.
This field of force builds the pressure
$$
p = {\chi\over2}  \ \Bv^{tot} \cdot \Bv^{tot}
= {\chi\over2} \ (\Bv^2+\Bv'^2) + \Mv\cdot\Bv'
$$
The first two terms are symmetrical about the box and exert no net force on it.
The last term gives
$$
\fv_2 = - \int (\Mv\cdot\Bv') \ d\Sv  
= - \int   (\Bv' \cdot d\Sv) \ \Mv 
= - g \, \Mv
\,.
$$
Here $d\Sv$ is the vector representation of a surface element
of the box and is directed outward.
Permuting $d\Sv$ and $\Mv$ was allowed because they are parallel 
in the region
where $\Bv'$ is important. The last equality comes from Gauss theorem
for magnetic charges.
Adding $\fv_1$ and $\fv_2$ one recovers the result (1)~:
$$
\fv = \fv_1 + \fv_2 = g \ (\Bv - \Mv) = g \, \Hv
\,.
\eqno(3)
$$
Most probably, (1) can be generalized to any shape of box.
Thus, the relevant field which drives a monopole in matter is $\Hv$. 
It is the field found in a parallel elongated cavity,
as for the force $\fv = e \, \Ev$ driving an electric charge inside a
dielectric.
{\it Mnemonic~:} this kind of cavity allows the test charge 
to follow the force without touching the matter. 

Eq.(1) allows trapping a (not too heavy) monopole in the pole 
of a permanent magnet, where the lines of $g\Hv$ 
converge from all directions. This would not be true for $g\Bv$.

\medskip
\noindent
{\it Microscopic interpretation}

Atomic magnetic dipoles are often pictured as microscopic
loops of electrical current. Then $\Bv$ appears as the 
average of the microscopic field $\bv$ over semi-macroscopic
volumes sufficiently large compared to the atomic scale. 
The work of $\bv$ along a straight line $L$ is therefore
$$
g \int_L \bv \cdot d\lv = g \int_L \Bv \cdot d\lv
\,.
$$
In contrast, the work along a line $L'$ which avoids passing through 
the loops is
$$
g \int_{L'} \bv \cdot d\lv = g \int_{L'} \Hv \cdot d\lv
\,.
$$
Eq.(1) implies that the monopole avoids passing through
the microscopic current loops or more likely that the loops move 
to "dodge" the monopole.
This was of course the case with the ferrofluid, 
but in solid matter the atoms cannot escape from the monopole trajectory.
Does it means that (1) is false if the monopole goes through
an atom~? Not necessarily. At the approach of the pole, the electron
wave functions are deformed and, if the monopole is sufficiently
slow, they return adiabatically to the ground states. 
Thus no energy is exchanged between the monopole and the atom, 
as if the loop "dodges" the monopole.

\medskip
\noindent
{\bf 3. VELOCITY-DEPENDENT FORCE}

In vacuum, the analog of the Lorentz force for a moving monopole is
$\fv = - g \, \vv \times \Ev$.
Accordingly, a piece $d\lv$ of wire carrying a current $I\dual$ 
of magnetic charges is subject to the dual Laplace force 
$d\fv = - I\dual \ d\lv \times \Ev$.
Following the ferrofluid example, we consider a wire protected 
by a waterproof tube in a liquid dielectric (Fig.2)~:

\begin{figure}
\centering\epsfig{file=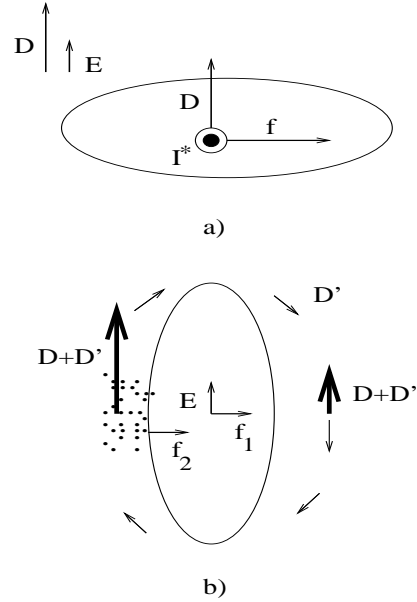,width=5.4cm,height=7.9cm}
\caption{Force acting on a wire carrying the dual current $I\dual$, 
in a dielectric liquid. The tube is flattened
(a) perpendicular (b) parallel to the field.}
\end{figure}

\noindent
a) the tube is flattened perpendicular to $\Dv = \varepsilon \Ev$. 
Then $\Ev_{tube}$ = $\Dv$
and the measured force is 
$$
d\fv = - I\dual \ d\lv \times \Dv
\,.
\eqno(4)
$$

\noindent
b) the tube is flattened parallel to $\Dv = \varepsilon \Ev$. 
Then $\Ev_{tube}$ = $\Ev$.
The force acting on the wire is $d\fv_1 = - I\dual \ d\lv \times \Ev$.
On the other hand, on the right of the tube, the total field 
$\Dv + \Dv'$ is larger than on the left.
The polar molecules are therefore attracted toward this region,
building an excess of pressure which pushes the tube toward the left.
Calculations like those between Eqs. (2) and (3) give 
the thrust $d\fv_2 = - I\dual \ d\lv \times \Pv$,
where $\Pv$ is the macroscopic electric polarization.
In total,
$$
d\fv = d\fv_1 + d\fv_2 = - I\dual \ d\lv \times (\Ev + \Pv)
$$
is equivalent to (4).
Thus the field acting on a wire of magnetic current is $\Dv$.
It is the field found in a perpendicular flattened cavity, 
as for the Laplace force $d\fv = I \ d\lv \times \Bv$ 
on an ordinary current in a magnetized matter.
{\it Mnemonic~:} this cavity allows the wire to follow the force 
without touching the matter.

For a moving monopole, (4) becomes 
$$
\fv = - g \ \vv \times \Dv
\,.
\eqno(4')
$$
 
\newpage

\noindent
{\it Microscopic interpretation}

In a dielectric, $\Ev$ is the average of the microscopic field $\ev$ 
over volumes sufficiently large compared to the molecular scale. 
The work of $\ev$ when the wire sweeps a flat surface $S$ is 
$$
I\dual \int\!\!\!\int_S \ev \cdot d\Sv 
= I\dual \int\!\!\!\int_S \Ev \cdot d\Sv 
\,.
$$
In contrast, the work of $\ev$ along a surface $S'$ which avoids cutting 
the dipole molecules is
$$
I\dual \int\!\!\!\int_{S'} \ev \cdot d\Sv 
= I\dual \int\!\!\!\int_{S'} \Dv \cdot d\Sv 
\,.
$$
Eq.(4) implies that a moving dual wire avoids cutting the dipole molecules, 
or that the molecules "dodge" the wire.
This has a well-defined topological meaning.
Let us recall however that this wire was introduced 
to make the problem time-independent.
For a moving monopole, there is no swept surface
and the topological interpretation is lost.

Gathering (1) and (4'), the total force on a magnetic charge is
$$
\fv = g \ (\Hv - \vv \times \Dv)
\,.
\eqno(5)
$$
This result does not take into account dissipation and holds only
for sufficiently slow monopoles, such that atoms
and molecules evolve adiabatically under the influence of the
monopole field.

\medskip
\noindent
{\bf 4. MAXWELL EQUATIONS IN DYONIC MATTER}

We consider matter containing magnetic charges $\pm g$
bound in magnetically neutral molecules, 
in addition to ordinary particles. These molecules posses magnetic dipoles 
of the form $g\rv$, building a macroscopic magnetic polarization $\pv\dual$
($\rv$ is the north-south charge separation). 
If they have spin, they also posses electric dipoles 
of the form $\gamma\antidual \Sv$ building a macroscopic electric 
polarization $\mv\antidual$ 
($\gamma\antidual$ is the "giro-electric" ratio). 
$\pv\dual$ and $\mv\antidual$ are dual respectively to the polarization $\pv$
and the magnetization $\mv$ built by the ordinary particles.
The dual-symmetric Maxwell equations for the space average of the 
microscopic fields are
$$
\rot \bar\bv - \partial_t \, \bar\ev = \jv+\delta\jv
$$
$$
\div \bar\ev = \rho + \delta\rho
$$
$$
- \rot \bar\ev - \partial_t \, \bar\bv = \jv\dual+\delta\jv\dual
$$
$$
\div \bar\ev = \rho\dual + \delta\rho\dual
\,.
$$
Here $(\rho, \jv)$ is the external ordinary charge-current 
density  and $(\delta\rho, \delta\jv)$ the induced one, given by
$$
\delta\rho = - \div \pv
$$
$$
\delta\jv = \rot \mv + \partial_t \, \pv
$$
Similarly, for the magnetic charge analogues, 
$$
\delta\rho\dual = - \div \pv\dual
$$
$$
\delta\jv\dual = - \rot \mv\antidual + \partial_t \, \pv\dual
$$
From these equations, we can write the dual-symmetric Maxwell equations
in matter~:
$$
\rot \Hv - \partial_t \, \Dv = \jv
$$
$$
\div \Dv = \rho
$$
$$
- \rot \Ev - \partial_t \, \Bv = \jv\dual
$$
$$
\div \Bv = \rho\dual
\eqno(6)
$$
where $\Hv = \bar\bv - \mv$, $\Dv = \bar\ev + \pv$ as usual,
but $\Ev = \bar\ev - \mv\antidual$ and $\Bv = \bv + \pv\dual$. 
We see that $\Ev$ and $\Bv$ can no more be interpreted 
as the spatial averages of the microscopic fields. In that
sense they are no more "fundamental" than $\Dv$ and $\Hv$.
In fact the dual of $\Ev$ is not $\Bv$ but $\Hv$,
whereas the dual of $\Bv$ is $-\Dv$.
The usual relations
$$
\Dv = \Ev + \Pv \,, \quad \Bv = \Hv + \Mv
\eqno(7)
$$
are recovered, defining 
$$
\Pv \equiv \pv + \mv\antidual \,,\quad \Mv \equiv \mv + \pv\dual
\,.
\eqno(8)
$$
It means that the microscopic nature of the dipole 
is forgotten at the level of the macroscopic Maxwell equations.
Only their long range fields in $1/r^3$ are relevant.
As in ordinary matter, $\Ev$ and $\Hv$ are the fields found 
in elongated cavities parallel to the respective fields, 
whereas $\Dv$ and $\Bv$ are found in perpendicular flat cavities. 

\newpage

\noindent
{\bf 5. THE DYONIC PERMITTIVITY-PERMEABILITY MATRIX}

We assume that the polarizations $\Pv$ and $\Mv$ respond linearly
to the macroscopic fields $\Dv$ and $\Bv$. 
$$
\pmatrix{\Pv \cr \Mv \cr} = [\chi]
\pmatrix{\Dv \cr \Bv \cr} = [\chi']
\pmatrix{\Ev \cr \Hv \cr}
\,,
\eqno(9)
$$
with $[1+\chi']\equiv[1-\chi]^{-1}$.
In ordinary matter $\chi_{11} = \chi_e = \chi'_e/\varepsilon$,
$\chi'_e \equiv \varepsilon-1$~; 
$\chi_{22} = \chi_m = \chi'_m/\mu$, 
$\chi'_m \equiv \mu-1$,
and $\chi_{12} = \chi_{21} = 0$.
In a matter containing only one species of dyon $(e,g)$ 
and antidyon $(-e,-g)$ bound in polar molecules,
$\Pv=\pv$ and $\Mv=\pv\dual$ are linked by
$$
{\pv \over e} = {\pv\dual \over g} 
\,, \quad 
{1\over e} \ {\partial p_i \over \partial D_j} 
= {1\over g} \ {\partial p_i \over \partial B_j}
\eqno(10)
$$
wherefrom 
$$
[\chi] = {\rm C}^{te} \ 
\pmatrix{e^2 & eg \cr eg & g^2\cr}
\,.
\eqno(11)
$$
An analogous matrix, with $e \leftrightarrow g$, is obtained  
with dipoles coming from spinning dyons ($\mv/e = \mv\antidual/g$).
We note that $[\chi]$ and $[\chi']$ are symmetrical matrices.
This remains true for a mixture of different species of molecules.

Thus, the usual relations $\Dv = \varepsilon \Ev$, $\Bv = \mu \Hv$
are replaced by
$$
\pmatrix{\Ev \cr \Hv \cr} = [1- \chi]
\pmatrix{\Dv \cr \Bv \cr}
\,,
\eqno(12)
$$
The speed of light is 
$$
c = (\det[1-\chi])^{1\over2} 
\quad
(c_{vac.} \equiv 1).
\eqno(13)
$$
Whatever they come from, the nondiagonal elements of $[\chi]$
violate P- and T- symmetries, since they connect vector to pseudovectors.
However PT is conserved.

\medskip
\noindent
{\bf 6. ENERGY-MOMENTUM TENSOR}

The various components of the energy-momentum tensor $\Theta^{\mu\nu}$ 
can be derived from energy and momentum conservation in simple physical systems.
Let us suppose that the whole space is filled with dual-symmetric matter.
To get $\Theta^{i0}$ (energy flow) and $\Theta^{ij}$ (momentum flow) 
one considers a sandwich made of three slab-like regions of the z coordinate,
R$_1$ = $[-a,0]$, R$_2$ = $[0,b]$ and R$_3$ = $[b,b+a]$.
R$_1$ carries uniform electric and magnetic charge-current densities, 
$\{\rho, \jv \;; \rho\dual, \jv\dual\}$ 
and R$_3$ carries the opposite densities, 
such that the fields vanish outside the sandwich.
Solving (6) and (12) with appropriate $\rho, \jv, \rho\dual, \jv\dual$, 
any kind of uniform field configuration $\{\Ev, \Dv \;; \Hv, \Bv\}$ 
can be obtained in R$_2$. 
These fields are linearly attenuated in R$_1$ and R$_3$. 
In R$_3$ a power 
$$
{dW\over dt\, d^3\rv} = \Ev\cdot\jv+\Hv\cdot\jv\dual
\eqno(14)
$$
is dissipated and a force 
$$
{d\fv\over d^3\rv} = \rho\Ev+\jv\times\Bv+\rho\dual\Hv-\jv\dual\times\Dv
\eqno(15)
$$
is exerted per unit of volume.
The same quantities per unit of area (integrated over $z$ in R$_3$) 
give $\Theta^{z0}$ and $\Theta^{zi}$ in R$_2$. 

To get $\Theta^{00}$ (energy density) and $\Theta^{0i}$ (momentum density)
one has to "rotate" the sandwich in the 4-dimensional space-time,
replacing $z$ by $t$ and slabs by time-slices or "epoch" T$_1$, T$_2$, T$_3$. 
During T$_1$ the (3-dimensional) space is filled
with uniform current densities $\jv$ and $\jv\dual$,
which progressively build uniform fields according (6) and (12).
The second epoch is current-free and the uniform fields remain constant.
The last epoch destroys the fields with opposite currents.
Integrating (14) and (15) over $t$ in T$_3$ 
give $\Theta^{00}$ and $\Theta^{0i}$ in T$_2$.
This method is detailled in [2]. 
One obtains 
$$
\Theta^{\mu\nu} = \pmatrix{
{\Ev\cdot\Dv + \Hv\cdot\Bv \over2} &
\Dv \times \Bv \cr
\Ev \times \Hv &
\Theta^{00} \delta^{ij} - D^i E^j - B^i H^j
\cr}
\eqno(16)
$$
as in ordinary matter. 

\smallskip
\noindent
{\it The Dirac condition in matter.}
One way to derive the Dirac condition between an electron and a monopole 
is to quantize the joint angular momentum of their fields which are 
$$
\Dv = {e \, \rv \over4\pi \, r^3}
\,,\quad
\Bv = {g \, \rv' \over4\pi \, r'^3}
$$
where $\rv$ (resp. $\rv'$) is the distance from the charge 
(resp. the pole) to the observation point.
According to (15) the momentum density is 
$\Theta^{0i} = (\Dv \times \Bv)_i$, 
from which one gets the angular momentum 
$$
\Jv = \int\!\!\!\int\!\!\!\int d^3\rv
\ \rv \times (\Dv \times \Bv)
= {eg \over 4\pi} \nv
\eqno(17)
$$
where $\nv$ is the unit vector from the charge toward the pole.
The usual Dirac condition
$eg = 2n\pi\, \hbar$
is obtained from the quantization rule
$\Jv \cdot \nv = n \hbar /2$.
Note that if the momentum density were $\Ev \times \Hv$,
as sometimes advocated (see the discussion in [3]),
the Dirac condition in medium would not be consistent
with that in vacuum.

\medskip
\noindent
{\bf 7. APPLICATION TO THE SEARCH FOR AN ELECTRIC DIPOLE MOMENT}

The dual-symmetric formalism applies as well to the case 
were the electron (or the nucleus) possesses 
an {\it electric dipole moment} (e.d.m.) $\vec d = \gamma\antidual \Sv$ 
in addition to the usual magnetic moment $\vec\magmom = \gamma \Sv$.
Then we have $\mv/\gamma = \mv\antidual/\gamma\antidual$
and a nondiagonal $[\chi]$ matrix element is generated, 
like with dyonic molecules (Eqs.10-11)~:
$$
[\chi] = 
\pmatrix{\chi_e + \chi_m r^2 & \chi_m r \cr
\chi_m r & \chi_m \cr}
\,,
\eqno(18)
$$
where $r\equiv\gamma\antidual/\gamma$.
$\chi_m = {\mu - 1 \over \mu}$ comes from the spinning electrons
and $\chi_e = {\varepsilon - 1 \over \varepsilon}$ from polar
molecules. 

A nonzero $\chi_{12}$ may be generated in another way~:
the e.d.m. tends to align the spin of an electron
along the internal electric field of a polar molecule.
It couples $\mv$ to $\pv$.
Here we consider only the first mechanism.

Eq.(18) suggests two possible measurements of $r$ :

\begin{figure}
\centering\epsfig{file=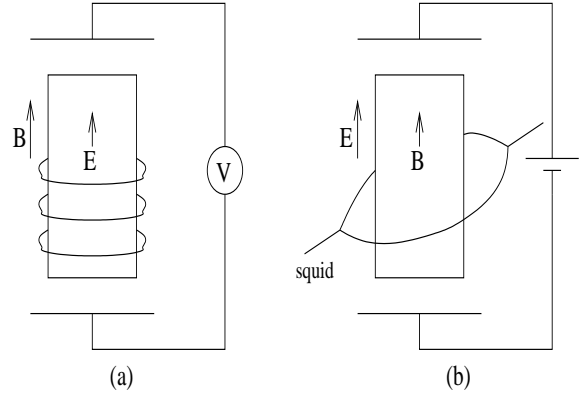,width=6.cm,height=8.1cm,angle=-90}
\caption{Scheme of e.d.m. search in macroscopic matter.
(a) container in a magnetic field~; a small potential difference
is measured with the voltmetre V.
(b) container in an electric field~; a small magnetic flux
is measured with a SQUID.}
\end{figure}

a) In Fig.3a, a cylinder of magnetizable, but insulating material 
is immersed in a large magnetic field $\Bv_0$. 
The inside field $\Bv$ induces a small electric polarization 
$\Pv = \chi_m r \Bv$ and an electric field $\Ev$.
If the cylinder is much broader than high, 
we have $\Bv=\Bv_0$, $\Dv \simeq 0$ and $\Ev \simeq - \Pv$. 
More generally one has
$$
\langle\Ev\rangle = - x \, \chi_m r \Bv_0
\eqno(19)
$$
where the coefficient $x<1$ depends on the container geometry.
Let us take a cubic container of size $L$. 
Between the top and the bottom, 
we can measure a potential difference $U = E L$.
The ratio between the stored electrostatic energy
$W = {1\over2} \varepsilon \Ev^2 L^3$  and the magnetic one
$W_0 = {1\over2\mu}  \Bv_0^2 L^3$ is 
$$
{W\over W_0} = \varepsilon\mu \,(x\chi_m r)^2
\,.
\eqno(20)
$$
in terms of common units,
we have 
\def\volt{\rm volt}
\def\metre{\rm metre}
\def\tesla{\rm tesla}
\def\joule{\rm joule}
\def\eV{\rm eV}
$$
{U\over\volt} = 3 \, 10^{8} 
\ x \, \chi_m \ r
{L\over\metre} \cdot {B_0\over\tesla}
$$
$$
{W\over\eV} = 2.5 \, 10^{24} 
\ x^2 \, \varepsilon \chi_m^2 \ r^2
\left({L\over\metre}\right)^3 \left({B_0\over\tesla}\right)^2
\,.
$$
[useful relations are~: 1 tesla = 3 10$^8$ volt/metre,

\noindent
1 (tesla)$^2$ (metre)$^3$ = 0.8 10$^6$ joule,

\noindent
1 eV = 1.6 10$^{-19}$ joule $\simeq 10^4$ kelvin
= $5 \, 10^6 \hbar$/metre

\noindent
and $\gamma_e=e\hbar/m_e = 4 \, 10^{-11}$ e$\times$cm].

Let us assume $r=10^{-16}$, which corresponds to
an electron e.d.m. of $2 \, 10^{-27}$ e$\times$cm,
$\varepsilon \sim1$, $\chi_m \sim 0.5$
and $x \sim 0.5$.
For a field of 1 tesla, and a cube of 1 meter,
a potential difference of about $0.5\,10^{-8}$ volt is obtained.
$W \sim 10^{-9}$ eV $\sim 10^{-5}$ kelvin.
The voltmetre has to be cooled at least to this temperature 
to prevent thermal noise.

b) in Fig.3b the same container is put in a large electric field $\Ev_0$.
Using the second matrix of (9)
$$
[\chi'] \simeq 
\pmatrix{\varepsilon - 1 & \varepsilon\chi'_m r \cr
\varepsilon\chi'_m r & \chi'_m \cr}
\,,
\eqno(18')
$$
with $\chi'_m\equiv \mu-1$,
we predict a small magnetization 
$\Mv = \varepsilon\chi'_m r \Ev$.
If the cylinder is much higher than broad, 
we have $\Ev \simeq \Ev_0$,
$\Hv \simeq 0$ and $\Bv \simeq \Mv$.
For a cubic container we assume
$$
\Bv \simeq x \, \varepsilon\chi'_m r \Ev_0
\eqno(21)
$$
with $x \sim 0.5$.
This field can be measured by a SQUID encircling the container.
The phase shift of the wave function in one loop is
$$
\varphi = e L^2 B /\hbar 
$$
$$
= 0.5 \, 10^{12} \ x \, \varepsilon\chi'_m \ r 
\left({L\over\metre}\right)^2 
{E_0 \over 10^5\, \volt/\metre}
\,.
$$
This phase can be multiplied by a large number of turns
around the cylinder.
The ratio between the output (magnetostatic) energy $W$
and the input (electric) one $W_0$ is still given by (20),
but $W$ and $W_0$ are typically $10^5$ times smaller 
and the temperature must be much lower than in case a).

\medskip
\noindent
{\bf 8. CONCLUSION}

We have given arguments that the macroscopic fields acting on 
magnetic charges and currents are $\Hv$ and $\Dv$. \
Comparing with electric charges and currents, 
one has a unified mnemonic principle~: 
in each case, the acting field
is the one found in a parallel-elongated (resp. flat-perpendicular) cavity 
in which a charge (resp. current wire) can follow the force
without touching the medium. 
In a classical microscopic picture,
a monopole avoids passing through the microscopic current loops
and a dual current wire avoids cutting the dipole molecules.
Quantum mechanically, it means 
that the perturbation of the atoms and molecules 
lying on the trajectory of the monopole is adiabatic.
This should be the case at low enough velocity in a liquid.
The monopole will be presumably accompanied by a swarm of atoms
magnetically (or electrically, for a dyon) bound to it.
In a solid, such a swarm could forbid the monopole to move 
without producing cracks.

The dual-symmetric Maxwell equations in matter are formally
unchanged, but $\Ev$ and $\Bv$ can no more be interpreted 
as the spatial averages of the microscopic fields.
The duality correspondance is
$\Ev \to \Hv$ and $\Dv \to \Bv$.
When dyons are present, or when ordinary particles
possess electric dipole moments,
$\varepsilon$ and $\mu$ are replaced
by a permittivity-permeability matrix $[1-\chi]$
whose nondiagonal elements violate the P- and T- symmetries (but not PT).
The energy momentum tensor is also unchanged.
The usual Dirac condition
$eg = 2n\pi\, \hbar$
is obtained provided the momentum density is 
$\Dv \times \Bv$.
These results have been obtained under the hypothesis
that $\Pv$ and $\Mv$ are linear in the fields.

As an application of the dual-symmetric formalism,
two possible measurements of the electron e.d.m. have been suggested. 
They are at the limit of the present technological
possibilities.
However, mechanisms like the $\mv - \pv$ coupling
in a polar molecule mentioned in Sect.7 might
enhance the signal.

\medskip

\centerline{\bf REFERENCES}

1. Y. Hara, Electromagnetic force on a magnetic monopole,
{\it Phys. Rev.}, 1985, v. A32, p.1002-1006

2. D. Fayolle, Dynamique d'un monop\^ole magn\'etique dans la mati\`ere,
{\it LYCEN}/9961, july 1999, unpublished.

3. F.N.H. Robinson, Electromagnetic stress and momentum in matter,
{\it Phys. Rep.}, 1975, v. C16, p. 313-354.

\end{document}